\documentclass[letterpaper,twocolumn,10pt]{article}
\usepackage{usenix-2020-09}

\usepackage{enumitem}
\usepackage{graphicx}
\usepackage{amsbsy}
\usepackage{stfloats}
\usepackage{multirow}
\usepackage[ruled]{algorithm2e}
\usepackage{tcolorbox}

\usepackage{tikz}
\usepackage{amsmath}
\usepackage{bm}
\usepackage{amssymb}

\usepackage{filecontents}

\begin{document}
%-------------------------------------------------------------------------------

%don't want date printed
\date{}

\title{\Large \bf DeMarking: A Defense for Network Flow Watermarking in Real-Time}

%for single author (just remove % characters)
\author{
    {\rm Yali Yuan}\\
    Southeast University
    \and
    {\rm Jian Ge}\\
    Southeast University
    \and
    {\rm Guang Cheng}\\
    Southeast University
}

\maketitle

%-------------------------------------------------------------------------------
\begin{abstract}
    %-------------------------------------------------------------------------------
    The network flow watermarking technique associates the two communicating parties
    by actively modifying certain characteristics of the stream generated by the sender
    so that it covertly carries some special marking information. Some curious users
    communicating with the hidden server as a Tor client may attempt de-anonymization
    attacks to uncover the real identity of the hidden server by using this technique.
    This compromises the privacy of the anonymized communication system. Therefore,
    we propose a defense scheme against flow watermarking. The scheme is based on
    deep neural networks and utilizes generative adversarial networks to convert
    the original Inter-Packet Delays (IPD) into new IPDs generated by the model.
    We also adopt the concept of adversarial attacks to ensure that the detector
    will produce an incorrect classification when detecting these new IPDs. This
    approach ensures that these IPDs are considered "clean", effectively covering
    the potential watermarks. This scheme is effective against time-based flow
    watermarking techniques.
\end{abstract}

%-------------------------------------------------------------------------------
\section{Introduction}
%-------------------------------------------------------------------------------

Traffic analysis is a technique for inferring sensitive information from network
traffic patterns, particularly packet timing and size, rather than packet content.
Because encryption does not significantly alter traffic patterns, traffic analysis
is primarily used in scenarios where network traffic is encrypted, particularly in
the dark web. Traditional traffic analysis, including website fingerprinting attacks
\cite{panchenko2016website,sirinam2019triplet,shen2023subverting} and flow correlation
attacks \cite{nasr2018deepcorr,oh2022deepcoffea,guan2023flowtracker},
primarily relies on passive analysis methods. These methods require the collection of a
significant amount of traffic data at key location nodes in the network, which involves
analyzing various traffic features to confirm the relationship between each network flow.
Hence, these processes demand substantial storage space and computational resources, making
it less scalable. As a result, active network flow watermarking technology has emerged.

Network Flow Watermarking (NFW) technology is primarily based on the concept of digital
watermarking \cite{cox2002digital}. NFW covertly embeds special marking information, known as watermarks,
into network flows by actively modifying certain characteristics of the flows generated
by the sender. After transmission through the communication network, if the receiver can
detect the corresponding watermark from the network flow, it is considered that there is
a network flow correlation between the sender and the receiver, and thus an explicit
communication relationship between them can be recognized. This technique offers the
advantages of reduced space-time overhead and high scalability. With the help of this
technique, some curious users (e.g., a law enforcement agency) can perform de-anonymization
attacks to uncover the real identity of the hidden server while communicating with it as
a Tor client, thereby compromising the privacy of the anonymous communication system.

In fact, after years of research, flow watermarking has a trend of evolving towards flow
fingerprinting \cite{houmansadr2013need}, which has the ability to embed more bits of information. Numerous
innovative schemes have been proposed. Through the use of various side channels, these
schemes achieve the dual effect of invisibility and robustness in attacks. Recently,
with the advancement of Deep Neural Network (DNN) technology, the first flow fingerprinting
scheme using DNN was proposed in the literature \cite{rezaei2021finn}. This scheme significantly enhances
the performance of flow fingerprinting and intensifies the threat to the privacy of anonymous
communication systems (e.g., Tor). Therefore, there is an urgent need for an effective defense
strategy, particularly against flow fingerprinting using DNNs.

Currently, the proposed schemes focus on detecting the presence of watermarks. For example,
Peng et al. \cite{peng2006secrecy} stated that the Kolmogorov-Smirnov (K-S) test can effectively detect
watermarking techniques based on the Inter-Packet Delay (IPD). Jia et al. \cite{jia2009blind} proposed
an MSAC attack method to detect the presence of DSSS watermarks. Defense against watermarking
does not seem to receive much attention.

Therefore, we propose a DNN-based defense scheme against flow watermarking for the first time.
This scheme utilizes Generative Adversarial Networks (GAN) \cite{goodfellow2014generative} to convert the original
Inter-Packet Delays (IPDs) into new IPDs generated by the model. We also draw inspiration
from adversarial attacks \cite{szegedy2013intriguing}, ensuring that the detector will produce an incorrect
classification when detecting these new IPDs. This approach guarantees that these IPDs
are "clean" and cover the potential watermarks. This scheme can effectively resist
time-based flow watermarking techniques. The main contributions of this paper are
as follows:

\begin{enumerate}[label=(\arabic*)]
    \item We propose a flow watermarking defense scheme named DeMarking which is based on adversarial
          attacks and GANs. The scheme can be utilized in the Tor network to enhance anonymity and
          provides a strong defense against time-based watermarking.
    \item We observe through extensive experiments that the small perturbations used in adversarial
          attacks in the image domain are unnecessary in the watermarking domain. From a defense
          standpoint, IPDs do not need to remain somewhat similar before and after the perturbation.
          Based on this property, we replace potentially watermarked IPDs with clean IPDs generated by
          the model, which significantly enhances the defense effect.
    \item We design a remapping function to control the range of the mean and standard deviation of
          the generated IPDs, thereby reducing the impact on the traffic rate and the performance of
          Tor. Additionally, we employ cosine similarity to transform the gradient boosting, commonly
          used in adversarial attacks, into gradient descent to facilitate the solution.
    \item We also experiment with conventional flow watermarking schemes that do not utilize
          deep neural networks. Our scheme also offers strong protection against conventional flow
          watermarking methods, such as RAINBOW \cite{houmansadr2009rainbow} and SWIRL \cite{houmansadr2011swirl}.
\end{enumerate}

The rest of this paper is organized as follows. In Section~\ref{sec2}, we investigate the
carrier of flow watermarking. In Section~\ref{sec3}, we introduce the scenario of this paper.
Sections \ref{sec4} and \ref{sec5} present some preliminaries and outline the scheme of
this paper. Sections \ref{sec6} and \ref{sec7} illustrate the experimental setups and
results. Finally, we discuss the limitations and future directions of our scheme in
Section \ref{sec8} and conclude the paper in Section \ref{sec9}.

%-------------------------------------------------------------------------------
\section{Related Works}
\label{sec2}
%-------------------------------------------------------------------------------

Since the concept of NFW was first proposed in 2001 \cite{wang2001sleepy}, researchers have explored numerous
carriers for embedding watermarks, including content, timing, size, and rate. The timing-based
carriers can be further subdivided into: IPD, Interval Centroid, and Interval Packet Counting.
Encrypted communication has become mainstream due to the widespread use of SSL/TLS protocols.
Content-based and size-based watermarking have gradually faded out of the researchers' focus
due to their unsuitability for encrypted traffic. Timing-based watermarking is widely welcomed
by researchers due to its better retention in network transmission. This is the reason why
there are many subcategories of timing-based carriers.

From the perspective of traditional telecommunications systems, all these carriers ultimately
fall into three diversity schemes: time diversity, frequency diversity, and space diversity.
Table \ref{table1} presents a collection of watermarking schemes and their corresponding carriers.

\begin{table}[]
    \centering
    \caption{Watermarking schemes and carriers.}
    \label{table1}
    \begin{tabular}{ccc}
        \hline
        \textbf{Watermarking Schemes}                  & \textbf{Diversities} & \textbf{Carriers} \\ \hline
        Wang et al. \cite{wang2001sleepy}              & -                    & Content           \\
        Wang et al. \cite{wang2003robust}              & Time                 & IPD               \\
        Wang et al. \cite{wang2007network}             & Time                 & Centroid          \\
        Yu et al. \cite{yu2007dsss}                    & Frequency            & Rate              \\
        Pyun et al. \cite{pyun2007tracing}             & Time                 & Counting          \\
        Houmansadr et al. \cite{houmansadr2009rainbow} & Time                 & IPD               \\
        Houmansadr et al. \cite{houmansadr2011swirl}   & Time                 & Centroid          \\
        Ling et al. \cite{ling2012novel}               & Time                 & Size              \\
        Iacovazzi et al. \cite{iacovazzi2017dropwat}   & Time                 & IPD               \\
        Rezaei et al. \cite{rezaei2021finn}            & Time                 & IPD               \\ \hline
    \end{tabular}
\end{table}

Through our investigation, we find that timing plays a significant role in watermarking
carriers. Even when certain watermarking schemes do not explicitly use timing as a carrier,
they are still influenced to some extent by time. For instance, when using rate as a carrier,
we can modify IPDs to change the transmission rate of the traffic.

Therefore, our scheme focuses on time-based watermarking as the primary adversary. We leverage
the concepts of GANs and adversarial attacks to transform the original IPDs in the network
transmission into clean IPDs generated by a model. This enables us to defend against time-based
watermarking effectively.

%-------------------------------------------------------------------------------
\section{Scenario}
\label{sec3}
%-------------------------------------------------------------------------------

An anonymous communication system refers to a communication system that employs techniques
such as data forwarding, content encryption, and traffic obfuscation on existing networks
to conceal communication content and relationships. Among them, Tor is one of the most widely
used anonymous communication systems. Our scenario is under the Tor network, which can be generally
categorized into three-hop and six-hop according to whether it adopts the hidden service technique
or not. Our proposed defense scheme, DeMarking, can be adapted to both scenarios. In this paper, we
choose the more challenging six-hop scenario to prove the efficiency of DeMarking.

\begin{figure}[htbp]
    \centering
    \includegraphics[width=\linewidth]{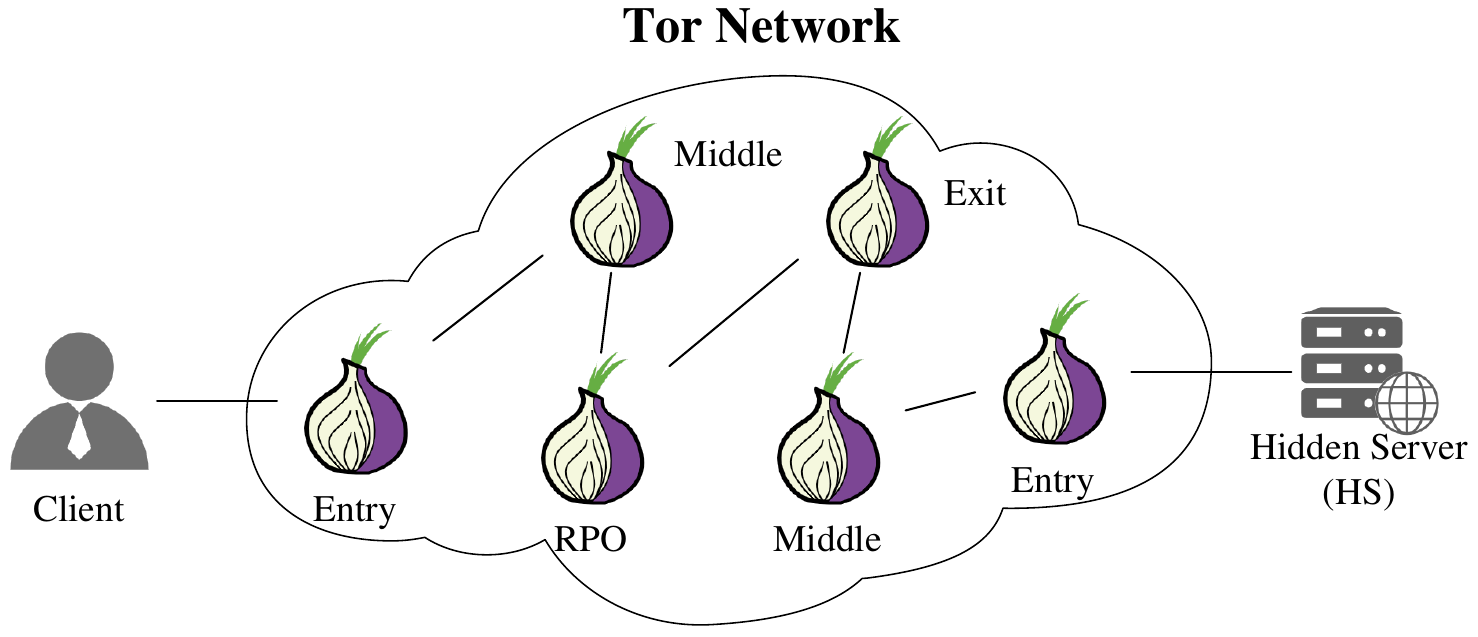}
    \caption{Tor network topology.}
    \label{fig1}
\end{figure}

The core idea of Tor is to provide protection for users' communication privacy through
multi-hop proxies and layered encryption. When a Tor Hidden Server (HS) is launched, it
selects three Tor nodes as its entry proxies. Similarly, when a client accesses the HS, it
also establishes three proxies to conceal its identity. After the initial configuration is
completed, the client selects a rendezvous point (RPO) as the convergence point for communication
between the client and the HS. Both the client and the HS establish links to the rendezvous point,
completing the construction of the 6-hop circuit, and communication can begin. Tor users access
hidden servers through a six-hop circuit, during which no single node can simultaneously know the
Tor client's IP address, the hidden server's IP address, and the data content. This ensures the
anonymity of the Tor client and the hidden server. The aforementioned process is illustrated
in Figure \ref{fig1}.

Network watermarking technology comprises two main components: the watermarker and the watermark
detector. The positions of these two components are designed and selected based on the desired
objectives and the expected observation points of the target traffic. In the Tor network scenario
depicted in Figure \ref{fig1}, the watermarker is typically located near the client to facilitate
embedding, while the watermark detector is generally positioned closer to the HS to determine its
true identity. The watermarker is responsible for converting watermark bits or fingerprint information
into watermark codes with specific characteristics and embedding them into the target flow. On the other
hand, the watermark detector monitors network traffic passing through specific nodes, analyzes traffic
features to detect watermarked flows, and subsequently decodes the watermark to extract the embedded
information.

\begin{figure}[htbp]
    \centering
    \includegraphics[width=\linewidth]{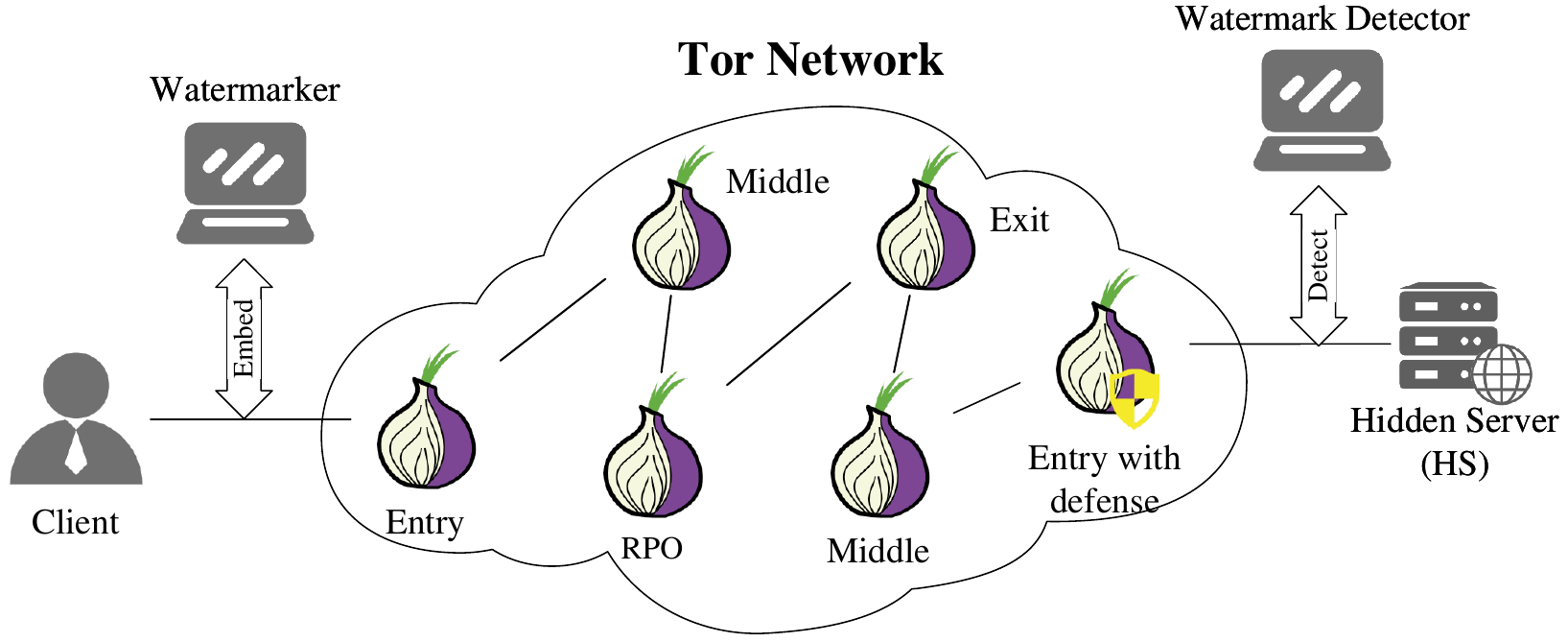}
    \caption{Flow watermarking and its defense.}
    \label{fig2}
\end{figure}

To defend against flow watermarking, the transformation of IPD must occur after watermark embedding
and before extraction. As depicted in Figure \ref{fig2}, the watermarker is located between the client
and its entry node, while the watermark detector is positioned between the HS and the entry node. In
this scenario, we can choose one or more of the six proxy nodes between the client and the HS to
deploy defense mechanisms. For simplicity, we select only the entry node of the HS as the Tor node
with defense.

%-------------------------------------------------------------------------------
\section{Preliminaries}
\label{sec4}
%-------------------------------------------------------------------------------

In this section, we introduce conventional watermarking techniques and DNN-based watermarking
techniques. Furthermore, we introduce adversarial attacks and migrate them to the watermarking
domain, laying the groundwork for our defense scheme.

\subsection{Flow Watermarking Techniques}

As stated in Section \ref{sec2}, time is a popular watermarking carrier. Therefore, we select two
state-of-the-art watermarking schemes based on different time carriers to introduce.

\textbf{RAINBOW:} Houmansadr et al. \cite{houmansadr2009rainbow} proposed the RAINBOW watermarking scheme, which is a
non-blind watermarking scheme based on IPD. In this scheme, the watermarker records the timestamps
of the passed packets and stores them in a database that is shared with the watermark detector.
Additionally, the watermarker introduces a certain delay in the IPD, which is also shared with
the watermark detector. When the traffic reaches the watermark detector, it records the timestamps
of the packets. By combining the IPD records in the database and calculating the normalized
correlation, the watermark detector can ascertain whether the flow has been marked.

\textbf{SWIRL:} The SWIRL watermarking scheme, also proposed by Houmansadr et al. \cite{houmansadr2011swirl},
is based on interval centroids. This scheme divides the time into a base interval and a mark
interval. The packets within the base interval are used to calculate the interval centroid,
which represents the average distance of the packets from the start of the interval. By quantizing
the interval centroid, the scheme determines the embedding mode used in the mark interval.
The mark interval is further subdivided into multiple sub-intervals, each of which is divided
into several time slots. The goal of embedding is to ensure that only the designated time slots
contain packets in each sub-interval, which can be achieved by introducing delays in the packets.
To detect the presence of a watermark, the detector also calculates the interval centroid based
on the base interval. Then, it measures the proportion of packets in the mark interval that fall
into the correct time slots. If this proportion exceeds a specific threshold, the watermark is
detected.

Our defense scheme provides good defense against time-based watermarking. In Section \ref{sec7}, we will
conduct an experimental analysis specifically targeting these two watermarking schemes.

\subsection{DNN-based Watermarking}
\label{sub4.2}

\begin{figure}[htbp]
    \centering
    \includegraphics[width=\linewidth]{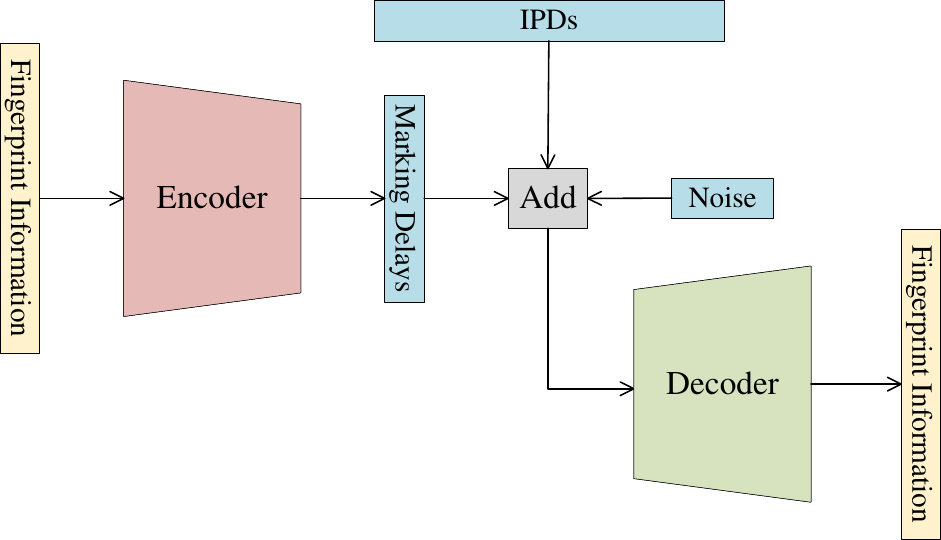}
    \caption{The architecture of DNN-based flow watermarking.}
    \label{fig3}
\end{figure}

DNN-based watermarking techniques often employ autoencoder architectures. Autoencoders were
initially proposed by Rumelhart et al. \cite{rumelhart1986learning} and have been widely applied in the field
of image watermarking \cite{fang2022pimog,liu2019novel,fu2022chartstamp,jia2021mbrs,fang2023denol}.
Recently, Rezaei et al. \cite{rezaei2021finn} applied autoencoders to the design
of flow fingerprinting schemes, using IPD as the watermark carrier to enhance the performance
of flow fingerprinting. The architecture of DNN-based flow watermarking is illustrated in
Figure \ref{fig3}.

As depicted in the figure, the input of the Encoder is a one-hot vector of length \textit{m},
representing the fingerprint information to be embedded. The output is a watermark delay
vector of length \textit{n}. Here, \textit{m} and \textit{n} are hyperparameters. To embed
the watermark, the watermark delay vector is added to a segment of IPD with the same length
of \textit{n}. Due to the inherent delays and jitter introduced during network transmission,
some noise is typically added to enhance the robustness of the model. The input of the Decoder
is the IPD with watermark delays and noise, which has the same length as the output of the
Encoder. The objective of the Decoder is to reconstruct the embedded fingerprint information,
which is the same as the input of the Encoder. The trained Encoder and Decoder are then deployed
as the watermarker and watermark detector, respectively, in the target network.

\subsection{Adversarial Attacks}
\label{sub4.3}

Adversarial attacks generally mislead the model to make incorrect classifications or predictions
by adding subtle perturbations to the model inputs. Such inputs to which subtle perturbations
are added are called adversarial samples. The generation of the adversarial sample $\boldsymbol{x}^*$
can be formulated as the following optimization problem:
\begin{equation}
    \boldsymbol{x}^{*}=\boldsymbol{x}+\arg \min _{\boldsymbol{\delta}} \forall \boldsymbol{x} \in D_{x}: O(\boldsymbol{x}+\boldsymbol{\delta}) \neq O(\boldsymbol{x}),
    \label{eq1}
\end{equation}
where $\boldsymbol{x}$ is the non-adversarial sample, $\boldsymbol{\delta}$ is the added
adversarial perturbation, $D_x$ represents the set of possible inputs, and $O(\cdot)$ denotes
the output of the target model or classifier. Since the target model $O(\cdot)$ is a deep
neural network, it is not possible to find a closed-form solution to this optimization
problem. Therefore, the equation can be numerically solved using empirical approximation
techniques.
\begin{equation}
    \arg \max _{\boldsymbol{\delta}} \mathbb{E}_{\boldsymbol{x} \in D_{x}}[L(O(\boldsymbol{x}+\boldsymbol{\delta}), O(\boldsymbol{x}))],
    \label{eq2}
\end{equation}
where $L(\cdot)$ is the loss function used to measure the distance between the outputs of
the target model before and after the perturbation.

The objective of an adversary is to add the minimal perturbation $\boldsymbol{\delta}$ to
force the target model to make incorrect classifications. Adversarial samples are commonly
studied in image classification applications. One constraint in finding adversarial samples
is that the added noise should be imperceptible to the human eye to ensure that the image
does not undergo significant changes.

Previous work \cite{nasr2021defeating} seemed to make this assumption when perturbing the IPD. It ensured that
there would be no significant changes in the IPD before and after perturbation. However,
when it comes to network communication, we are not concerned about the exact shape of the
IPD. Our main concern is to ensure that it does not have a significant impact on the
transmission rate. Under this condition, we can completely use an IPD sequence $\boldsymbol{y}$
that has a similar mean and standard deviation to $\boldsymbol{x}$ but is entirely different from it.

To achieve this, we propose a converter model C. When provided with the original IPD sequence
$\boldsymbol{x}$, the converter transforms it into a new IPD sequence $\boldsymbol{y}=C(\boldsymbol{x})$.
Therefore, our optimization objective becomes optimizing the parameters of the converter model C.
Thus, equation \ref{eq2} can be adjusted as follows:
\begin{equation}
    \arg \max _{C} \mathbb{E}_{\boldsymbol{x} \in D_{x}}[L(O(C(\boldsymbol{x})), O(\boldsymbol{x}))].
    \label{eq3}
\end{equation}

\subsection{Cosine Similarity}
\label{sub4.4}

Previous works \cite{goodfellow2014explaining,moosavi2017universal,dong2018boosting} have utilized gradient boosting to maximize the loss of the target
model in adversarial attacks. To enhance implementation efficiency, we employ cosine similarity
as the distance metric, which transforms the gradient boosting into gradient descent. The
formula for calculating cosine similarity is as follows:
\begin{equation}
    \cos(\boldsymbol{a},\boldsymbol{b})=\frac{\boldsymbol{a}\cdot \boldsymbol{b}}{|\boldsymbol{a}|\cdot|\boldsymbol{b}|}=\frac{\sum_{i=1}^na_ib_i}{\sqrt{\sum_{i=1}^na_i^2}\cdot\sqrt{\sum_{i=1}^nb_i^2}}.
    \label{eq4}
\end{equation}

The cosine similarity is a value between -1 and 1. When vectors $\boldsymbol{a}$ and $\boldsymbol{b}$
are more similar, $cos(\boldsymbol{a},\boldsymbol{b})$ approaches 1, and conversely, it approaches -1.
Therefore, the equation \ref{eq3} can be finally expressed as follows:
\begin{equation}
    \arg\min_C\mathbb{E}_{\boldsymbol{x}\in D_x}\left[\cos\left(O(C(\boldsymbol{x})),O(\boldsymbol{x})\right)\right].
    \label{eq5}
\end{equation}

%-------------------------------------------------------------------------------
\section{Design of DeMarking}
\label{sec5}
%-------------------------------------------------------------------------------

\begin{figure*}[ht]
    \centering
    \includegraphics[width=0.9\linewidth]{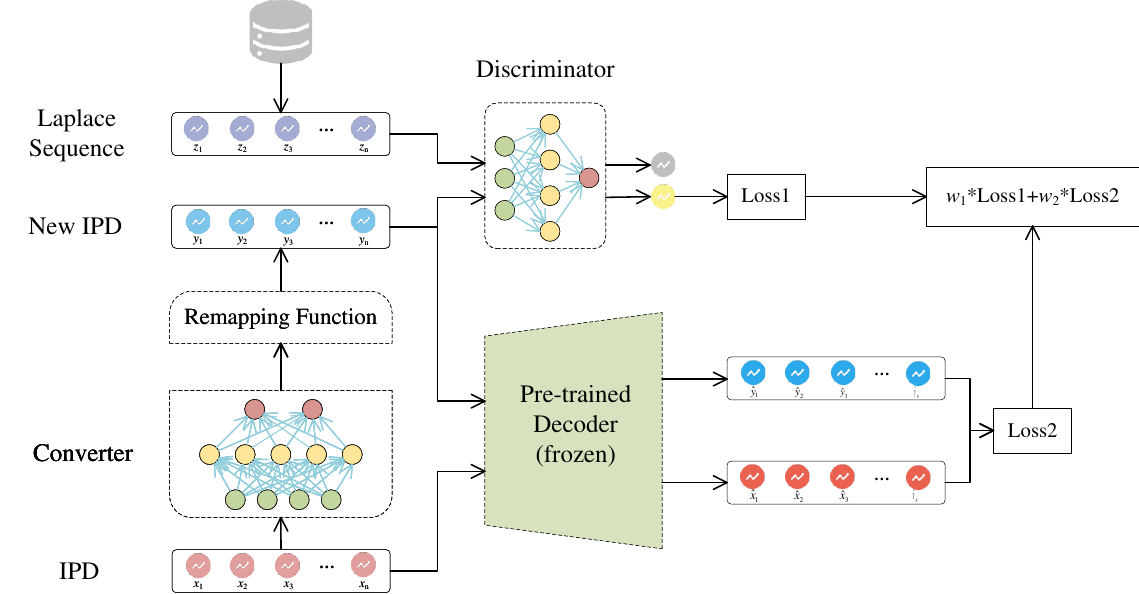}
    \caption{Training architecture.}
    \label{fig4}
\end{figure*}

Since the IPD sequence should adhere to the expected statistical distribution of the target
protocol, we design a discriminator based on the concept of GAN to guide the generation of IPDs
by the converter. In conjunction with the adversarial attack discussed in Section \ref{sub4.3},
we design the following architecture to train the converter to generate IPD sequences that meet
the desired criteria, as shown in Figure \ref{fig4}.

\textbf{Converter:} The converter is a fully connected network whose parameters are learned
during training. Its input is an IPD sequence $\boldsymbol{x}=\{x_1,x_2,\cdots,x_n\}$ of length
\textit{n}. The output is a new IPD sequence $\boldsymbol{y}=\{y_1,y_2,\cdots,y_n\}$ of the same
length, obtained after passing through a remapping function. This new IPD sequence
$\boldsymbol{y}$ will be clean and will replace the original IPD to implement the defense mechanism.

\textbf{Discriminator:} The discriminator is also a fully connected network. It is responsible
for distinguishing between the data generated by the converter and the data from the target
distribution (e.g., Laplace distribution), guiding the converter to generate data that conforms
to a specific distribution. The input of the discriminator, similar to the converter, is a vector
of length \textit{n}. The output of the discriminator is transformed into a probability value through
the sigmoid function.

\textbf{Decoder:} The decoder is a pre-trained model discussed in Section \ref{sub4.2}, and its
parameters are frozen. Under the white-box condition, it has the same structure and parameters as
the adversary's decoder. Under the black-box condition, it is a custom structure trained as
described in Section \ref{sub4.2}. The decoder extracts the embedded fingerprint information
of length \textit{m} from the IPD sequences of length \textit{n}. In the figure \ref{fig4},
$\boldsymbol{x}$ and $\boldsymbol{y}$ are input to the decoder, and $\widehat{\boldsymbol{x}}$
and $\widehat{\boldsymbol{y}}$ are obtained as outputs.

Since we employ a generative adversarial network, we have two loss functions to control the
performance of each task: the discriminator loss and the converter loss. For the discriminator,
we utilize the Mean Absolute Error (MAE) to evaluate its discrimination effectiveness. As for
the converter, we aim to generate data that conforms to a specific distribution while also
misleading the decoder through adversarial attacks. To balance these objectives, we introduce
two weights, $w_1$ and $w_2$, resulting in $w_1*Loss1+w_2*Loss2$ as depicted in the figure \ref{fig4}.
Here, $Loss1$ represents the scores provided by the discriminator, and $Loss2$ measures the
distance between the extracted vectors using the cosine similarity described in Section \ref{sub4.4}.

Table \ref{table2} provides a detailed overview of the structures of the converter and the discriminator.
The Adam optimizer is employed for training in both cases.

\begin{table*}[htb]
    \centering
    \caption{Details of the model structure.}
    \label{table2}
    \begin{tabular}{c|cc}
        \hline
                                   & Layer             & Details                           \\ \hline
        \multirow{3}{*}{Converter} & Fully Connected 1 & Size: 1024, Activation: LeakyRelu \\
                                   & Fully Connected 2 & Size: 2048, Activation: LeakyRelu \\
                                   & Fully Connected 3 & Size: 512, Activation: LeakyRelu  \\ \hline
        Discriminator              & Fully Connected 1 & Size: 2048, Activation: Relu      \\ \hline
    \end{tabular}
\end{table*}

In addition to ensuring that the IPD sequence conforms to a specific distribution, it is also
important to minimize the introduction of excessive delay during defense, as this can interfere
with the underlying application. Therefore, we design a remapping function $R(\cdot)$ to control
the value of the generated IPD sequence. The specific details are as follows:
\begin{equation}
    \begin{array}{c}
        R\left(C(\boldsymbol{x}), \mu_{\min }, \mu_{\max }, \sigma\right)=C(\boldsymbol{x}) * \frac{\min (\operatorname{std}(C(\boldsymbol{x})), \sigma)}{\operatorname{std}(C(\boldsymbol{x}))}- \\
        \max \left(\overline{C(\boldsymbol{x})}-\mu_{\max }, 0\right)-\min \left(\overline{C(\boldsymbol{x})}-\mu_{\min }, 0\right),
    \end{array}
    \label{eq6}
\end{equation}
where $\overline{C(\boldsymbol{x})}$ represents the mean of the generated new IPD sequence
$C(\boldsymbol{x})$, $\mu_{\min }$ and $\mu_{\max }$ are the lower and upper bounds, respectively,
that we aim to restrict the range of $\overline{C(\boldsymbol{x})}$, and $\sigma$ is the maximum
allowable standard deviation. By utilizing this remapping function, we can control the mean of
the newly generated IPD sequence within a specific range.

To apply the trained converter to real-time network traffic, we design Algorithm \ref{alg1}.
During the defense initiation, a random sequence of length \textit{n},
$\boldsymbol{x_0}=\{x_{0,1},x_{0,2},\cdots,x_{0,n}\}$, is generated. $\boldsymbol{x_0}$ is
then fed into the converter and passed through the remapping function, resulting in a new
IPD sequence, $\boldsymbol{y_0}=\{y_{0,1},y_{0,2},\cdots,y_{0,n}\}$. When network packets arrive,
their timestamps, $\boldsymbol{t_0}=\{t_0,t_1,\cdots,t_n\}$, are recorded, and the IPD is
replaced with $\boldsymbol{y_0}$. Once $\boldsymbol{y_0}$ is exhausted,
$\boldsymbol{x_1}=\{t_1-t_0,t_2-t_1,\cdots,t_n-t_{n-1}\}$ is computed based on the timestamps
$\boldsymbol{t_0}$. The subsequent defense process is repeated following the aforementioned steps.

\begin{algorithm}[!h]
    \caption{DeMarking defense scheme}
    \label{alg1}
    \DontPrintSemicolon
    \SetKwFor{While}{while}{:}{}
    \SetKwFor{For}{for}{\string:}{}
    \SetKwIF{If}{ElseIf}{Else}{if}{:}{elif}{else:}{}

    \SetKwInOut{Data}{Data}
    \Data{ \;
    $n \gets$ the length of the IPD sequence\;
    $\boldsymbol{q} \gets$ the packet buffer queue\;
    $\boldsymbol{t} \gets$ the packet timestamp buffer queue\;
    $\mu_{\min } \gets$ lower bound on the mean value of the generated IPDs\;
    $\mu_{\max } \gets$ upper bound on the mean of the generated IPDs\;
    $\sigma \gets$ upper bound on the standard deviation of the generated IPDs\;
    $\boldsymbol{x}=\{x_{0,1},x_{0,2},\cdots,x_{0,n}\} \gets$ randomly initialized IPDs\;
    $\boldsymbol{y}=R(C(\boldsymbol{x}),\mu_{\min},\mu_{\max},\sigma) \gets$ Convert $\boldsymbol{x}$ to obtain new IPDs
    }
    \BlankLine
    \begin{minipage}{\linewidth}
        \textbf{Thread 1:}\;
        \nlset{1}\While{a packet arrives}{
            \nlset{2}put the packet into $\boldsymbol{q}$\;
            \nlset{3}record the timestamp into $\boldsymbol{t}$\;
        }
        \BlankLine
    \end{minipage}
    \BlankLine
    \begin{minipage}{\linewidth}
        \textbf{Thread 2:}\;
        \nlset{1}\While{true}{
        \nlset{2}\While{$\boldsymbol{y}$ is not empty}{
            \nlset{3}sleep($y_0$) and remove $y_0$ from $\boldsymbol{y}$\;
            \nlset{4}\If{$\boldsymbol{q}$ is not empty}{
                \nlset{5}send packet $q_0$ and remove $q_0$ from $\boldsymbol{q}$\;
            }
            \nlset{6}\Else{
                \nlset{7}send a random packet\;
            }
        }
        \nlset{8}$l = min(n, len(\boldsymbol{t}))$\;
        \nlset{9}\For{$i\in\{1,2,\cdots,l\}$}{
        \nlset{10}$x_i=t_i-t_{i-1}$\;
        }
        \nlset{11}\If{$l<n$}{
            \nlset{12}\For{$i\in \{l+1,\cdots,n\}$}{
                \nlset{13}$x_i =$ a random IPD\;
            }
        }
        \nlset{14}remove $t_0,t_1,\cdots,t_l$ from $\boldsymbol{t}$\;
        \nlset{15}$\boldsymbol{y}=R(C(\boldsymbol{x}),\mu_{\min},\mu_{\max},\sigma)$
        }
    \end{minipage}

\end{algorithm}

%-------------------------------------------------------------------------------
\section{Experimental Setup}
\label{sec6}
%-------------------------------------------------------------------------------

In this section, we discuss our experimental setup. All our code was written in Python,
and the DNN techniques were implemented using PyTorch \cite{paszke2019pytorch}.

\subsection{Metrics}

Flow watermarking and flow fingerprinting techniques differ in terms of the number of
embedded bits. Watermarking techniques only consider embedding a single watermark bit,
while flow fingerprinting embedding consists of multiple bits representing fingerprint
information. Therefore, for these two techniques, we employ different evaluation metrics.

For watermarking techniques, we employ the True Positive (TP) and False Positive (FP) rates
as evaluation metrics.
\begin{itemize}
    \item TP: The ratio of flows correctly identified as containing watermarks and actually
          embedded with watermarks, to the total number of flows with watermarks.
    \item FP: The ratio of flows incorrectly identified as containing watermarks but not actually
          embedded with watermarks, to the total number of flows without watermarks.
\end{itemize}

For flow fingerprinting techniques, we utilize the Extraction Rate (ER) and Bit Error Rate (BER)
as evaluation metrics.
\begin{itemize}
    \item ER: The ratio of the number of flows from which the fingerprint can be successfully
          extracted to the total number of flows with fingerprints.
    \item BER: We convert each fingerprint to its binary representation to calculate the BER.
          It is important to note that we consider the average bit error rate across all fingerprints.
\end{itemize}

\subsection{Adversary Setup and Models}
\label{sub6.2}

We assume that the flow fingerprinting adversary adopts the model structure described in
reference \cite{rezaei2021finn} and utilizes the Adam optimizer. The specific details are
shown in Table \ref{table3}.

\begin{table*}[htbp]
    \centering
    \caption{Details of the adversary model.}
    \label{table3}
    \begin{tabular}{c|cc}
        \hline
                                 & Layer                                & Details                      \\ \hline
        \multirow{4}{*}{Encoder} & Fully Connected 1                    & Size: 1000, Activation: Relu \\
                                 & Fully Connected 2                    & Size: 2000, Activation: Relu \\
                                 & Fully Connected 3                    & Size: 2000, Activation: Relu \\
                                 & Fully Connected 4                    & Size: 500, Activation: Relu  \\ \hline
        \multirow{9}{*}{Decoder} & \multirow{4}{*}{Convolution Layer 1} & Kernel number: 50            \\
                                 &                                      & Kernel size: 10              \\
                                 &                                      & Stride: (1,1)                \\
                                 &                                      & Activation: Relu             \\ \cline{2-3}
                                 & \multirow{4}{*}{Convolution Layer 2} & Kernel number: 10            \\
                                 &                                      & Kernel size: 10              \\
                                 &                                      & Stride: (1,1)                \\
                                 &                                      & Activation: Relu             \\ \cline{2-3}
                                 & Fully Connected 1                    & Size: 128, Activation: Relu  \\ \hline
    \end{tabular}
\end{table*}

In the white-box scenario, we can directly utilize the known model and its parameters.
However, in the black-box scenario, we lack knowledge about the adversary's model, parameters,
or dataset. Therefore, we leverage the transferability of adversarial attacks
\cite{papernot2016transferability,papernot2017practical} to
design a custom substitute model. Based on this substitute model, we train a converter to
defend against the adversary's model.

In order to demonstrate the effectiveness of our approach against traditional watermarking
techniques as well, we reproduce RAINBOW \cite{houmansadr2009rainbow} and SWIRL \cite{houmansadr2011swirl}
using Python. We conduct experiments on these techniques using the trained converter.

\subsection{Datasets}

To obtain real-world traffic data, we set up a traffic collection environment based on the
scenario described in Section \ref{sec3}. We use two Alibaba Cloud servers running Ubuntu 20.04,
with Tor version 0.4.2.7 installed. One server is located in Hong Kong, serving as the client,
while the other server is located in Singapore, serving as the HS. The client accesses the HS
through six real-world Tor nodes. Since watermarking techniques only consider the one-way link
from the client to the HS, we generate traffic by uploading a randomly generated 3MB file. We
divide the experiments into two groups, with each group performing 500 uploads to obtain datasets
$D_1$ and $D_2$, respectively. Due to the automatic circuit switching feature of Tor, our data
is not always collected on the same circuit. Based on the source IP, destination IP, and protocol
type, we can easily distinguish the target flows and extract the packet timestamps for calculating IPD.

%-------------------------------------------------------------------------------
\section{Experimental Results}
\label{sec7}
%-------------------------------------------------------------------------------

To demonstrate the effectiveness of our scheme, we first conduct defense experiments on
the adversary model described in Section \ref{sub6.2} under both white-box and black-box conditions.
Subsequently, we perform defense experiments on RAINBOW and SWIRL using the trained converter.
Finally, we evaluate the computational cost of our approach.

\subsection{White Box}
\label{sub7.1}

Assuming that we want to embed fingerprint information of 10 bits, the input to the Encoder
is a one-hot vector of length 1024. For the output of the Encoder, we select four different
output lengths: 50, 100, 150, and 200, corresponding to four different model sizes. We set
the remapping function to control the mean of the generated IPDs between 30 and 60, with a
standard deviation of less than 20. We use dataset $D_1$, with 80\% of the data used for training
and 20\% for testing.

First, we train the adversary model on the training set until convergence. Then, we use the
IPD sequences with watermarking delays generated by the adversary model as input to the Converter
and train it based on the training architecture shown in Figure \ref{fig4}.

We measure the extraction rate and bit error rate of the model with and without defense.
The results are shown in Table \ref{table4}.

\begin{table*}[ht]
    \centering
    \caption{ER and BER with and without defense in white-box scenarios.}
    \label{table4}
    \begin{tabular}{c|cc|cc}
        \hline
        \multirow{2}{*}{Length of IPD sequences} & \multicolumn{2}{c|}{ER}                  & \multicolumn{2}{c}{BER}                                                               \\ \cline{2-5}
                                                 & \multicolumn{1}{c|}{Without defense(\%)} & With defense(\%)        & \multicolumn{1}{c|}{Without defense(\%)} & With defense(\%) \\ \hline
        50                                       & \multicolumn{1}{c|}{99.307}              & 0.098                   & \multicolumn{1}{c|}{0.316}               & 50               \\ \hline
        100                                      & \multicolumn{1}{c|}{97.542}              & 0.098                   & \multicolumn{1}{c|}{1.290}               & 50               \\ \hline
        150                                      & \multicolumn{1}{c|}{99.211}              & 0.098                   & \multicolumn{1}{c|}{0.403}               & 50               \\ \hline
        200                                      & \multicolumn{1}{c|}{98.809}              & 0.098                   & \multicolumn{1}{c|}{0.587}               & 50               \\ \hline
    \end{tabular}
\end{table*}

From Table \ref{table4}, we can observe that our defense scheme is highly effective.
Without defense, the adversary model can successfully extract the fingerprint information
with extraction rates of over 95\% and low bit error rates. However, with defense,
the extraction rate of the adversary model is nearly zero, and the bit error rate increases
to 50\%. This aligns with our calculated expectation: given a bit sequence of length $l$,
randomly generating another bit sequence of the same length, we can compute the expected
bit error rate as $\frac{1*C_l^1+2*C_l^2+\cdots+l*C_l^l}{2^{l}*l}=0.5$.

\subsection{Black Box}

In the black-box scenario, we assume that the adversary uses a model with an output
length of 100, as described in Section \ref{sub7.1}. Since we do not have access to the
details of the adversary's model or the dataset they use, we first establish a substitute
model with an input length of 1024. The details of the substitute model are shown in
Table \ref{table5}.

\begin{table*}[htbp]
    \centering
    \caption{Details of the substitute model.}
    \label{table5}
    \begin{tabular}{c|cc}
        \hline
                                 & Layer             & Details                      \\ \hline
        \multirow{2}{*}{Encoder} & Fully Connected 1 & Size: 500, Activation: Relu  \\
                                 & Fully Connected 2 & Size: 2000, Activation: Relu \\ \hline
        \multirow{2}{*}{Decoder} & Fully Connected 1 & Size: 1000, Activation: Relu \\
                                 & Fully Connected 2 & Size: 3000, Activation: Relu \\ \hline
    \end{tabular}
\end{table*}

It can be observed that our substitute model is simple yet effective. We also select output
lengths of 50, 100, 150, and 200 for the substitute model. To differentiate it from the
adversary's dataset, we train the substitute models of different sizes on the dataset $D_2$
until convergence. Based on these substitute models and the training architecture shown in
Figure \ref{fig4}, we train the corresponding Converters. It is important to note that
the training of the Converters is also conducted on dataset $D_2$.

After the training is completed, we perform defense experiments against the adversary model
on the dataset $D_1$ using Converters with lengths of 50, 100, 150, and 200, respectively.
Dataset $D_1$ is the dataset used by the adversary model. The defense results are presented in
Table \ref{table6}.

\begin{table}[htbp]
    \centering
    \caption{ER and BER with defense in black-box scenarios.}
    \label{table6}
    \begin{tabular}{ccc}
        \hline
        Length of IPD sequences & ER(\%) & BER(\%) \\ \hline
        50                      & 0.033  & 49.997  \\
        100                     & 0.167  & 49.833  \\
        150                     & 0.133  & 50.407  \\
        200                     & 0.133  & 49.620  \\ \hline
    \end{tabular}
\end{table}

It is evident that without defense, the extraction rate and bit error rate of the adversary
model are 97.542\% and 1.290\% respectively, as shown in Table \ref{table4}. After applying
defense in the black-box scenario, the extraction rate is nearly reduced to zero, and the bit
error rate increases to around 50\%, as shown in Table \ref{table6}. Therefore, our defense
scheme demonstrates effective results in both white-box and black-box scenarios.

Unlike conventional schemes that just add perturbations to IPDs, our scheme involves a
comprehensive transformation of IPDs. The magnitude of the changes introduced is significantly
higher compared to perturbations, leading to alterations in IPD patterns. Consequently,
after the defense, it becomes challenging for the Decoder to extract the fingerprint information.

\subsection{Conventional Watermarking Techniques}

In order to demonstrate the effectiveness of our defense against conventional watermarking
techniques, we conduct experiments on RAINBOW and SWIRL using the four Converters trained
in the black-box scenario. The experiments are conducted using the dataset $D_1$, which differs
from the dataset $D_2$ used for training the models.

\begin{figure}[htbp]
    \centering
    \includegraphics[width=0.8\linewidth]{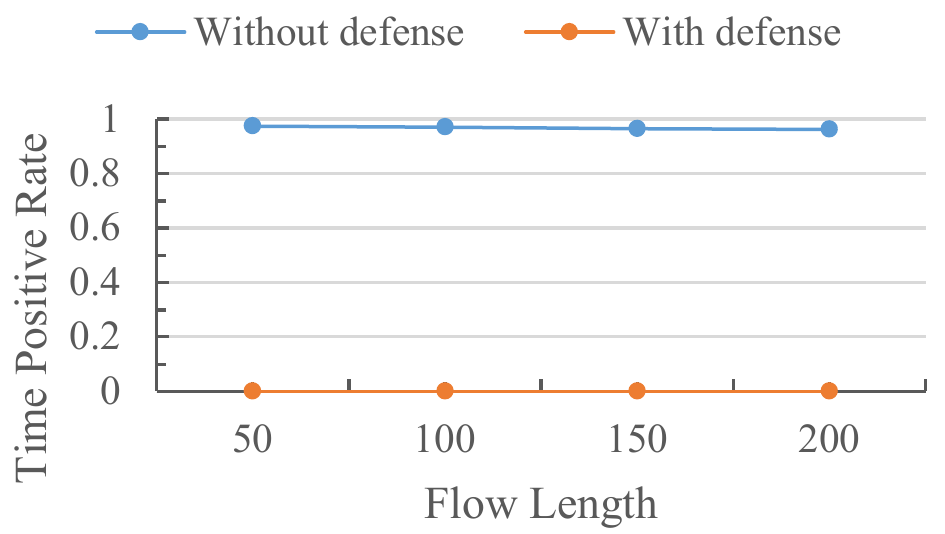}
    \caption{Experimental results of RAINBOW.}
    \label{fig5}
\end{figure}

\textbf{RAINBOW:} We use the first 1200 IPDs of each flow in the dataset $D_1$ for watermark
embedding. The watermark perturbation amplitude is set to 10, and the detection threshold
for the watermark is set to 0.54. These parameter values are recommended in the literature.
During the experiments, we simulate network jitter using a Laplace distribution with a location
parameter of 0 and a scale parameter of 10. The changes in the TP rate with and without defense
are shown in Figure \ref{fig5}.

It can be observed that after applying our defense, the TP rate of RAINBOW is always
reduced to zero regardless of the flow length, which shows the robustness of the DeMarking.
We also observe that the FP rate is close to zero regardless of whether the defense is applied
or not, so we do not list it.

\textbf{SWIRL:} SWIRL is based on interval centroids and is more complex than RAINBOW.
The parameters for SWIRL are also set to the recommended values found in the literature.
The interval length is set to 2 seconds, with 20 subintervals and 5 slots. The packet
detection threshold is set to 0.5, and the mark detection threshold is set to 12 out of 32
interval pairs. In the absence of defense, our experiments show that SWIRL achieves a TP rate
of 96.689\%. However, after applying our defense, regardless of which Converter is used, the
TP rate is reduced to 0\%. Similar to RAINBOW, the FP rate is nearly zero regardless of the
presence or absence of defense.

\subsection{Computational Costs}

\begin{figure}[htbp]
    \centering
    \includegraphics[width=\linewidth]{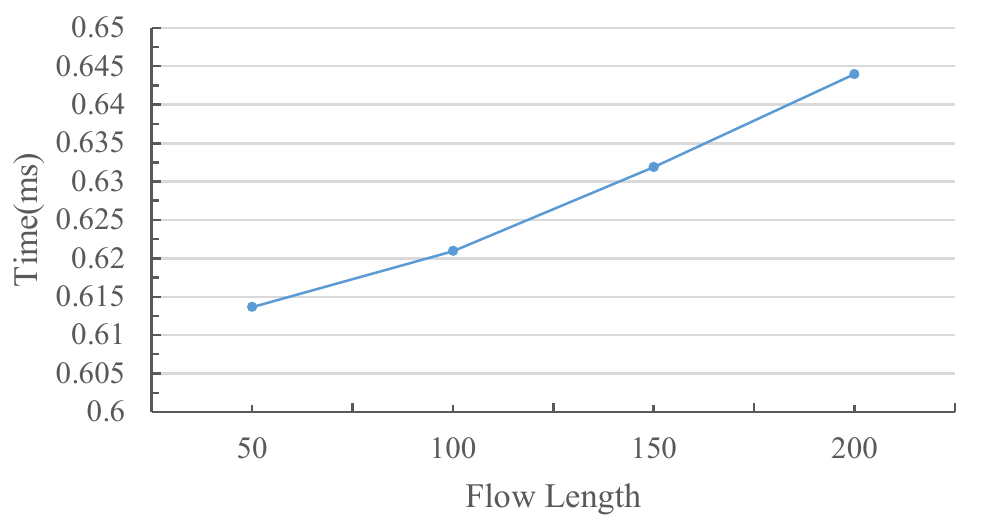}
    \caption{Computational costs of converters.}
    \label{fig6}
\end{figure}

Since the defense of watermarking needs to be performed in real-time over network transmissions,
the computational cost of the defense components is an important metric for evaluating the
feasibility of the defense. The GPU we use is GeForce RTX 3090. To evaluate the computational
cost, we measure the conversion elapsed time of one IPD record as an indicator. We use one IPD
record as the input to the Converter and calculate the average time over 10,000 iterations.
The measured conversion time is shown in Figure \ref{fig6}.

It can be observed that the conversion time for a single IPD sequence is approximately 0.6
milliseconds, which is negligible for IPDs. Therefore, our defense approach is capable of
handling real-time network transmissions.

%-------------------------------------------------------------------------------
\section{Limitations and Future Directions}
\label{sec8}
%-------------------------------------------------------------------------------

As previously mentioned, this work focuses on defending against time-based watermarking.
Our study shows that our proposed defense scheme achieves good results against flow
fingerprinting\cite{rezaei2021finn} based on DNN and some conventional watermarking
techniques \cite{houmansadr2009rainbow,houmansadr2011swirl}.
To accomplish this, we incorporate techniques such as GANs and adversarial attacks.

However, it is important to emphasize that our defense scheme is only applicable to
watermarking schemes influenced by IPDs. For other types of watermarks based on content,
size, rate, and other carriers, our defense method may not be applicable. Therefore,
in future research, we encourage the inclusion of a wider range of watermark carriers
within our defense framework to enhance the defense capabilities against various types
of watermarking techniques.

It should be noted that the inclusion of these watermark carriers may require further
research and exploration. We believe that by continuously improving and expanding the
defense against watermarking, we can provide more effective protection for our communication
privacy.

%-------------------------------------------------------------------------------
\section{Conclusions}
\label{sec9}
%-------------------------------------------------------------------------------

In this paper, we introduce a defense scheme against time-based watermarking which can be
called DeMarking. DeMarking utilizes techniques such as GAN and adversarial attacks to
transform the original IPD sequence into a new, clean IPD sequence, thereby achieving
defense. We also construct a remapping function to minimize the impact on network flows.

We evaluate our defense against DNN-based flow fingerprinting techniques. Even in the black-box
scenario, our defense demonstrates good effectiveness. We also evaluate our defense against
conventional watermarking techniques, including different time carriers, and achieve satisfactory
defense results. Finally, we measure the computational cost of our scheme. The computational
time required for our defense is negligible compared to an IPD, allowing it to handle real-time
network transmissions.

%-------------------------------------------------------------------------------
\bibliographystyle{plain}
\bibliography{\jobname}

%%%%%%%%%%%%%%%%%%%%%%%%%%%%%%%%%%%%%%%%%%%%%%%%%%%%%%%%%%%%%%%%%%%%%%%%%%%%%%%%
\end{document}